\documentclass[a4paper,11pt]{article}
\pdfoutput=1 % if your are submitting a pdflatex (i.e. if you have
\usepackage{jinstpub} % for details on the use of the package, please
\usepackage{multirow}

\title{The NEWS-G detector at SNOLAB}
\collaboration{NEWS-G collaboration}

\author[m]{L.~Balogh,}
\author[g]{C.~Beaufort,}
\author[a,1]{A.~Brossard,\note{Corresponding author.}}
\author[m]{J.-F.~Caron,}
\author[a]{M.~Chapellier,}
\author[a]{J.-M.~Coquillat,}
\author[d]{E. C.~Corcoran,}
\author[a]{S.~Crawford,}
\author[g]{A.~Dastgheibi-Fard,}
\author[i]{Y.~Deng,}
\author[a]{K.~Dering,}
\author[i]{D.~Durnford,}
\author[i]{C.~Garrah,}
\author[a]{G.~Gerbier,}
\author[c]{I.~Giomataris,}
\author[a]{G.~Giroux,}
\author[f]{P.~Gorel,}
\author[c]{M.~Gros,}
\author[a]{P.~Gros,}
\author[g]{O.~Guillaudin,}
\author[b]{E.~W. Hoppe,}
\author[j]{I.~Katsioulas,}
\author[d]{F.~Kelly,}
\author[j]{P.~Knights,}
\author[f]{S.~Langrock,}
\author[l]{P.~Lautridou,}
\author[j]{I.~Manthos,}
\author[a]{R.D.~Martin,}
\author[j]{J.~Matthews,}
\author[c]{J.-P.~Mols,}
\author[g]{J.-F.~Muraz,}
\author[j]{T.~Neep,}
\author[j]{K.~Nikolopoulos,}
\author[i]{P.~O'Brien,}
\author[i]{M.-C.~Piro,}
\author[a]{N.~Rowe,}
\author[g]{D.~Santos,}
\author[d]{P.~Samuleev,}
\author[a]{G.~Savvidis,}
\author[h]{I.~Savvidis,}
\author[l]{F.~Vazquez~de~Sola~Fernandez,}
\author[a]{M.~Vidal,}
\author[j]{R.~Ward,}
\author[g]{M.~Zampaolo.}

\affiliation[a]{Department of Physics, Engineering Physics \& Astronomy, Queen’s University, Kingston, Ontario, K7L 3N6, Canada}
\affiliation[b]{Pacific Northwest National Laboratory, Richland, Washington 99354, USA}
\affiliation[c]{IRFU, CEA, Université Paris-Saclay, F-91191 Gif-sur-Yvette, France}
\affiliation[d]{Chemistry \& Chemical Engineering Department, Royal Military College of Canada, Kingston, Ontario K7K 7B4, Canada}
\affiliation[f]{SNOLAB, Lively, Ontario, P3Y 1N2, Canada}
\affiliation[g]{LPSC, Universit\'{e} Grenoble-Alpes, CNRS-IN2P3, Grenoble, 38026, France}
\affiliation[h]{Aristotle University of Thessaloniki, Thessaloniki, 54124 Greece}
\affiliation[i]{Department of Physics, University of Alberta, Edmonton, T6G 2E1, Canada}
\affiliation[j]{School of Physics and Astronomy, University of Birmingham, Birmingham, B15 2TT, United Kingdom}
\affiliation[k]{Department of Physics and Astronomy, Laurentian University, Sudbury, Ontario, P3E 2C6, Canada}
\affiliation[l]{SUBATECH, IMT-Atlantique/CNRS-IN2P3/Nantes University, Nantes, 44307, France}
\affiliation[m]{Department of Mechanical and Materials Engineering, Queen’s University, Kingston, Ontario K7L 3N6, Canada}

\emailAdd{alexis.brossard@queensu.ca}

\abstract{
The New Experiments With Spheres-Gas (NEWS-G) collaboration intends to achieve sub-$\mathrm{GeV/c^{2}}$ Weakly Interacting Massive Particles (WIMPs) detection using Spherical Proportional Counters (SPCs). SPCs are gaseous detectors relying on ionisation with a single ionization electron energy threshold. The latest generation of SPC for direct dark matter searches has been installed at SNOLAB in Canada in 2021. This article details the different processes involved in the fabrication of the NEWS-G experiment. Also outlined in this paper are the mitigation strategies, measurements of radioactivity of the different components, and estimations of induced background event rates that were used to quantify and address detector backgrounds. }


\begin{document}
\maketitle
\flushbottom

\section{Introduction}
\label{sec:intro}
Astrophysical observations, from galactic to cosmological scales, show that a significant part of the universe is made of invisible or dark matter~\cite{Planck:2018vyg, Thompson:2014zra, Clowe:2006eq}. None of the fundamental particles in the standard model of particle physics nor a combination of them, are able to constitute more than a small fraction of dark matter~\cite{Billard:2021uyg}, however,
%have characteristics that agree with the properties of this this invisible matter. %Thus, identifying a particle responsible for this unknown mass is a challenge for both our understanding of the Universe and the incomplete standard model of particle physics. 
several models for physics beyond the standard model give viable candidate particles~\cite{Feng:2010gw}.
%that could solve the dark matter issue. 
%, each in a mass range lying between $\mathrm{10^{-5}}$ and $\mathrm{10^{7}}$\,$\mathrm{eV/c^{2}}$~\cite{Calmet:2020pub}. 
Well motivated candidates include Weakly Interacting Massive Particles (WIMPs), such as those provided by super symmetric extensions to the Standard Model~\cite{Catena:2013pka}, as well as axions~\cite{Peccei:1977hh}. Furthermore, recent theoretical models with sub-GeV candidate masses have motivated the expansion of searches to include light dark matter~\cite{Essig:2022dfa}.   

The NEWS-G collaboration is searching for light dark matter candidates (<1\,$\mathrm{GeV/c^{2}}$) using Spherical Proportional Counters (SPCs)~\cite{Giomataris_2008}. SPCs are spherical vessels filled with gas, a design which offers several advantages for light mass dark matter searches. The combination of a high gain and low intrinsic electronic noise allows for the detection of single ionization electrons~\cite{Gerbier:2014jwa}. This low energy threshold, in addition to the use of targets with low atomic mass, such as H or Ne, increases the sensitivity to sub-$\mathrm{GeV/c^{2}}$ dark matter candidates. In 2015, the NEWS-G collaboration set a new limit for WIMP masses lower than 0.5\,$\mathrm{GeV/c^{2}} $ at a cross-section of $\mathrm{4.4 \times 10^{-37}\,cm^{2}}$  with the 60 cm SEDINE detector. The detector was filled with a neon and methane mixture at 3.1 bar and operated at the Laboratoire Souterrain de Modane (LSM)~\cite{Gerbier:2004ad} in France for 41 days of data acquisition~\cite{Arnaud:2017bjh}. 

Since then, the collaboration has developed a new 135 cm in diameter detector. This detector, larger than the previous generation, is made of materials stringently selected for their radio-purity. In addition, a new multi-anode sensor, "ACHINOS" (Fig~\ref{fig:ACHINOS}), was developed to ensure a sufficiently strong electric field in this larger volume, while retaining high gain capabilities~\cite{Giganon:2017isb,Giomataris:2020rna}. After a commissioning run at the LSM in 2019, the detector has been installed at SNOLAB in Canada~\cite{Smith:2012fq}. 

In this article we present the fabrication of this detector. Section~\ref{sec:SPC} focuses on the detector operating principle, and gives a description of the sensor and the data acquisition system. 
Section~\ref{sec:NEWS-G} presents the details concerning the fabrication of the detector and its low radioactivity materials shielding. 
Section~\ref{sec:BKGs} details the different strategies used for background mitigation. 
Section~\ref{sec:Activity} outlines measurements of the different activity components of the experiment, and details expectations regarding background event rates.  
This article presents the detector construction and shows the expected main source of background and is a support for dedicated data analysis paper to be published.

% ************************************** %
\section{The Spherical Proportional Counter}
\label{sec:SPC}
% ************************************** %
The SPC was invented at the Comissariat à l'Energie Atomique (CEA) in France by Ioannis Giomataris~\cite{Giomataris_2008}. The detector is made of a spherical vessel filled with gas
and an anode in its center. The anode is biased to a positive high voltage (HV) of up to a few thousand volts via an insulated HV wire going through the grounded support rod. 
The advantages of this technology include:
\begin{itemize}
  \item The simplicity of the detector and the possible use of copper (the metal with the highest radiopurity potential) as the main constituent.
  \item The very low capacitance of the sensor, combined with a high-gain operation of the detector, allows for an energy threshold low enough to observe single electron ionization.
  \item Few channels are required to read-out a large volume and large target mass at high pressure.
  \item A time projection chamber capability with multi-ball anode.
  \item Capability of identifying volume events by pulse shape analysis providing a strong background rejection.
  %\item Capability of discrimination of particles, within the rise-time range of volume interactions by the analysis of the shape of the pulse. This provides a strong background rejection capability.
\end{itemize}

SPCs have been implemented for several purposes. A first detector, SEDINE, was developed for dark matter search and installed at the LSM. The R2D2 collaboration is developing an SPC for neutrinoless double beta decay searches~\cite{Bouet:2020lbp}. This technology can also be applied in neutron spectroscopy~\cite{Bougamont:2015jzx}. As a consequence of its simplicity, it is also useful for teaching and outreach events.

\subsection{Operating Principle}
\label{sec:Principle}
\begin{figure}[h!]
  \centering
  \includegraphics[width=0.45\linewidth]{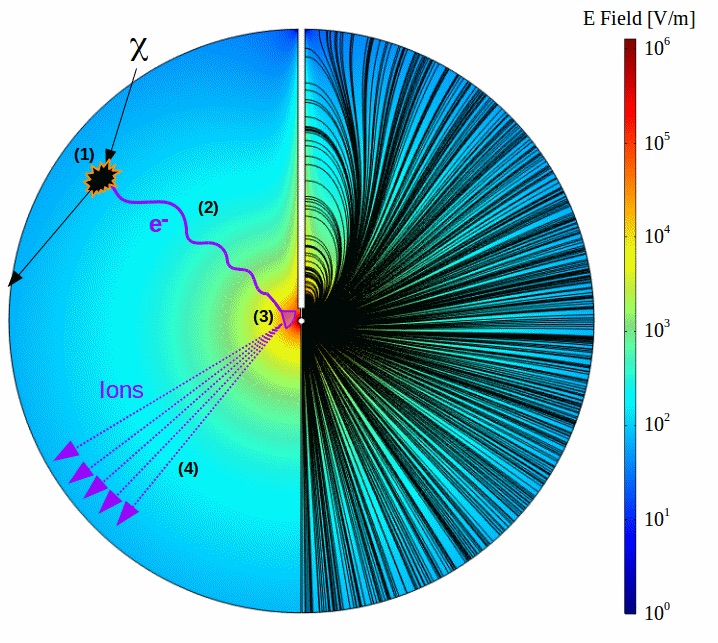}
  \caption[Functioning of an SPC]{Schematic view of the spherical proportional counter and detection principle. The various stage of the signal formation are depicted: (1) Ionization, (2) primary electrons drift, (3) charge avalanche, and (4) avalanche ions drift. The black lines show the electric field lines.}
  \label{fig:SchemaSPC}
\end{figure}

The operating principle of the detector is illustrated in Fig~\ref{fig:SchemaSPC}. (1) When particles interact in the gas volume, they ionize gas molecules, inducing the emission of primary electrons. 
%The number of primary electrons produced is proportional to the energy deposited by the particle and is dependent on the gas properties, including ionization potential and the mean energy required to create an ion electron pair (W-value). 
(2) Under the influence of the electric field, the primary electrons drift toward the anode at the center of the sphere. The diffusion of the primary electrons in a large drift volume allows for the resolution of individual electrons in time. Typical drift times are around a few hundreds of $\mathrm{\mu s}$, depending on gas pressure and composition. (3) When primary electrons reach the amplification region,
within approximately 100\,$\mathrm{\mu m}$ of the anode,
the strong electric field provides sufficient kinetic energy to produce secondary ion-electron pairs, creating a charge avalanche. (4) The majority of the current is  induced by the motion of the avalanche ions toward the cathode. The current is then integrated through a charge amplifier and digitized.

\subsection{The ACHINOS Sensor}
\label{sec:ACHINOS}
\begin{figure}[h!]
\centering
\includegraphics[width=0.4\linewidth]{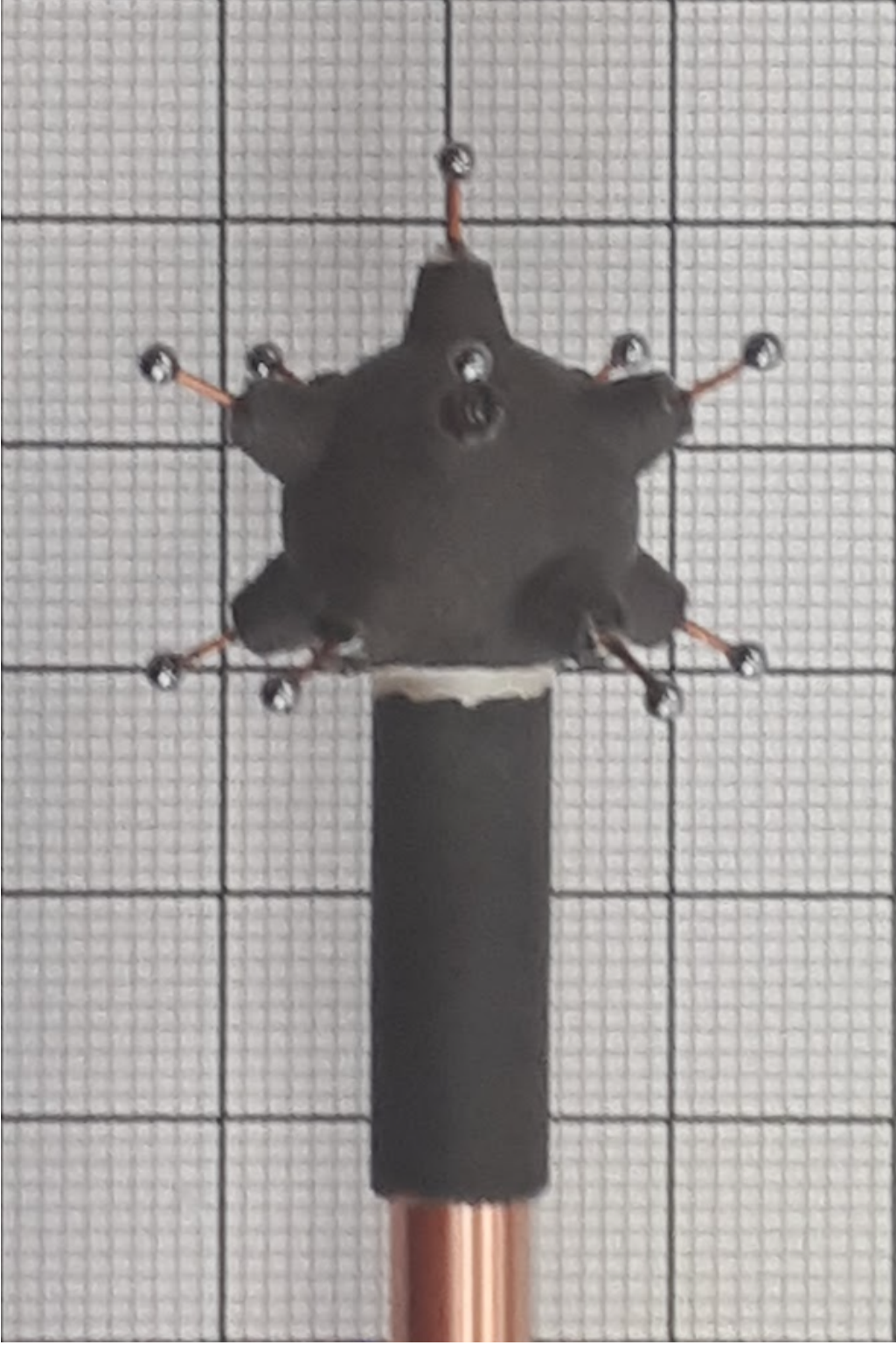}
\caption[ACHINOS sensor]{An 11-anode ACHINOS sensor.}
\label{fig:ACHINOS}
\end{figure}

\begin{figure}[h!]
\centering
\includegraphics[width=0.7\linewidth]{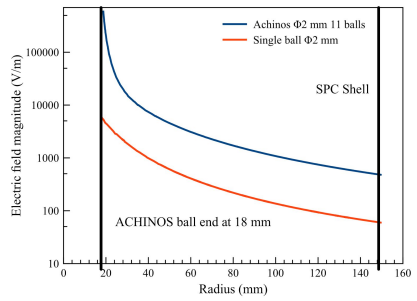}
\caption[ACHINOSF]{Comparison of the electric field strength for single 2 mm anode ball (orange) and a 36 mm diameter Achinos with 11, 2 mm anode balls (blue)}
\label{fig:ACHINOSF}
\end{figure}

The electric field for a single anode as a function of radius is given by
\begin{equation}
    E(r)=\frac{V}{r^{2}} \left ( \frac{r_{S}r_{A}}{r_{S}-r_{A}} \right )\,,
\end{equation}
where $V$ is the voltage applied to the anode, $r$ is the radial distance from the anode, $r_{A}$ is the radius of the anode, and $r_{S}$ the radius of the sphere. As $r_{S}\gg r_{A}$ the electric field can be approximated by
\begin{equation}
    E(r)\approx \frac{V}{r^{2}}  {r_{A}}\,,
\end{equation}
thus, the electric field is dictated mainly by the anode radius and voltage. For large detector operated at higher pressure $P$, the smaller ratio $E/P$, results in an increased probability of primary electron attachment or recombination. This can be overcome by increasing the anode radius and voltage but this would increase the chance of detector instabilities and discharges.      
%In the drift region the strength of the electric field is proportional to the radius of the anode. Thus, a large anode ensures a strong electric field throughout the volume of the detector, allowing a high charge collection efficiency. However, close to the anode, $r \approx r_{A}$, the electric field is inversely proportional to the radius of the anode. 
%A small anode ensures a strong amplification field and a large gain. For large volume detectors, the drift field becomes too weak for efficient operation. 
%To avoid electron attachment and recombination, a strong drift field is needed, requiring a large anode. On the other hand, the high gain needed for a single ionization electron energy threshold demands a small anode. To overcome these contradictory requirements, 
A new sensor concept, ACHINOS, has been developed to overcome these issues~\cite{Giganon:2017isb,Giomataris:2020rna}. ACHINOS, shown in Fig~\ref{fig:ACHINOS}, comprises several anodes at a fixed radius around a central support structure.
%The wires have a constant length and are fixed perpendicularly to the surface of a central sphere in such a way that all the individual anode balls are located on a virtual sphere. 
ACHINOS decouples the drift and avalanche electric fields, with the drift field being determined by the collective electric field of all anodes, and the avalanche field being determined by the individual anode to which the electron arrives. This enables the detector to operate with the desired avalanche gains, while still obtaining the sufficient drift electric field magnitude. This permits operation of larger detectors and higher pressures without compromising detector stability. Fig~\ref{fig:ACHINOSF} shows a comparison between a single anode ball and an ACHINOS.
The central support structure also acts as an electrode, which is used to tune the electric field. The central electrode is made of either highly resistive material or a total insulator covered by a resistive surface, such as a Diamond-Like Carbon (DLC) coating, making it possible to apply a secondary voltage to the surface~\cite{Katsioulas:2018pyh}.
%The advantage of this technique is that at large distance, the electric field is enhanced, in first approximation by a factor close to the number of anode balls. Near each individual anode, the amplification field is inversely proportional to their radius. 
%In addition by reading out the individual anode the detector acquires TPC-like characteristics with 3D track reconstruction.

\subsection{Electronics and Data Acquisition Systems}
\label{sec:DAQ}
The charge-sensitive preamplifier integrates the input current to give an output voltage signal proportional to the number of electric charges collected by the anode. The pulse produced decreases with a time constant $\mathrm{\tau}$ characteristic of each amplifier. Charge sensitive preamplifiers were selected for their gain and time constant. The CREMAT CR-110 charge sensitive pre-amplifier was used, which has a decay time constant of $\mathrm{\tau=140\,\mu s}$ and gain of $\mathrm{1.4\,V/pC}$. 
The amplified signal is digitized by a "CALI" box digitizer produced at the CEA in Saclay~\cite{dastg2014}. It functions in a dynamic range of $\mathrm{\pm 1.25\,V}$ with a maximum sampling rate of 5 MHz. Different internal gain settings are available, including $\times1$, $\times1.333$, $\times2$ or $\times4$.
The full digitized signal is sent to a computer through an Ethernet cable. The data acquisition and triggering is handled by the SAMBA~\cite{dastg2014} program developed at the CEA Saclay and used by the EDELWEISS~\cite{Armengaud:2017rzu}, CUPID-MO~\cite{Armengaud:2019loe} and R2D2~\cite{Bouet:2020lbp} experiments. SAMBA is used to handle the full digitized data flow produced by the CALI digitizer. It performs user-defined optimized triggering to select and store the useful data on disk. SAMBA also provides the user with a visual display of the data, useful for diagnostic and monitoring purposes.

% ************************************** %
\section{The NEWS-G detector at SNOLAB}
\label{sec:NEWS-G}
\begin{figure}[h!]
\centering
\includegraphics[width=0.45\linewidth]{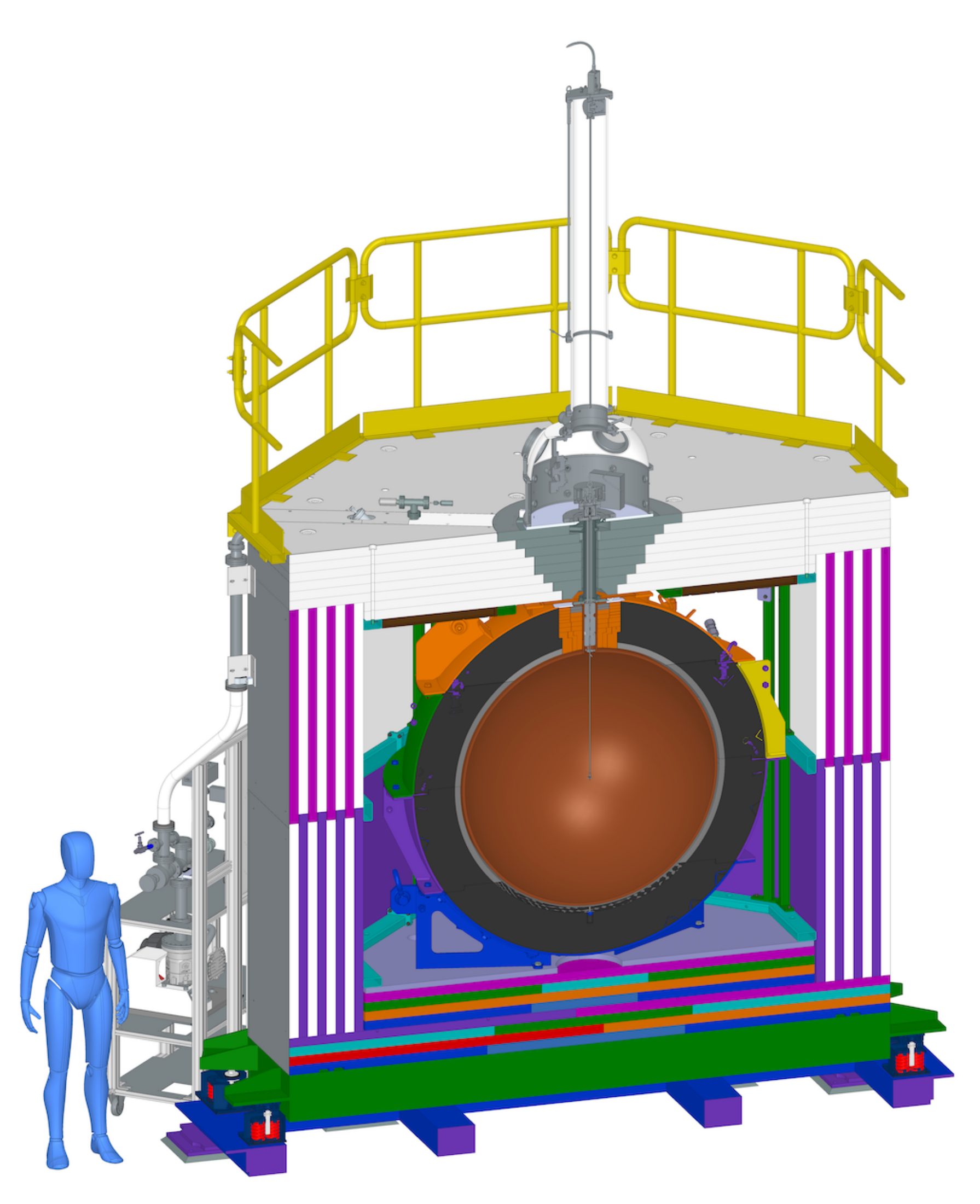}
\caption[Schematic view of the experiment. ]{Schematic view of the experiment. The spherical copper vessel (orange) and the lead shielding (black) are enclosed in a stainless steel shell and HDPE walls. The gas and vacuum piping, and electronics connect to the sphere on the top. The rod and sensor anode can be installed through a glove box on the top. The detector and its shielding sit on a seismic platform (green)}
\label{fig:Schematic}
\end{figure}

% ************************************** %
The detector developed by the collaboration is presented in Fig~\ref{fig:Schematic}. It is made of a 135 cm diameter spherical copper vessel enclosed in a lead and polyethylene shield. The whole shield is mounted on a seismic platform to prevent damage from seismic events in the mine where SNOLAB is located.

\subsection{The 135-cm Diameter Spherical Copper Vessel}
\label{sec:Sphere}

The spherical vessel is made of 99.99\% pure C10100 Oxygen-Free Electronic Copper. Initially, two disks were formed and stored at the LSM. Both disks were spun to form two hemispheres. A 500\,µm thick layer of pure copper was then deposited on the inner surface of the hemispheres by electroplating~\cite{Balogh:2020nmo} to mitigate the background induced by the $\mathrm{^{210}Pb}$ contamination of the commercial copper. Simulations of the plating show a reduction of the $\mathrm{^{210}Pb}$ induced background by more than one order of magnitude (Fig~\ref{fig:Plating_Decay}).  This amount was decided taking into account background mitigation requirement and time constraint, the plating of the hemispheres took 20 and 22 days. The two hemispheres were finally welded together using electron beam welding. High-pressure certification tests were conducted, and identified leaks were corrected using Tungsten Inert Gas (TIG) welding. Following the welding, the inner and outer surfaces of the vessel were etched using a solution of hydrogen peroxide and sulphuric acid~\cite{Bunker:2020sxw}. An estimated 5\,µm of copper was removed from the inside of the vessel, carrying away its surface contamination from exposure to air. The inner surface was then passivated with citric acid~\cite{Hoppe}.
A second etching of the inner surface with hydrogen peroxide and sulphuric acid was performed under a
nitrogen atmosphere at SNOLAB in 2020,
removing about 8\,µm of copper. 

\subsection{The Shielding}

%\subsubsection{The lead shielding}
%\label{sec:Pb}
The copper vessel is surrounded by a 25\,cm thick spherical lead shield against gamma background consisting of
a 3\,cm thick inner layer of Roman lead followed by a 22\,cm thick layer of commercial low activity lead. The Roman lead shell allow for shielding the background coming from the $\mathrm{^{210}Pb}$ decay chain within the modern lead layer. 
The two layers are shown in light and dark grey, respectively, in Figure~\ref{fig:Schematic}. The shield consists of six large pieces, each weighing 4 to 5\, tonnes. A hole to insert the rod and gas piping is filled with smaller pieces made of Roman lead (orange in Figure~\ref{fig:Schematic}).
The lead is encased in an airtight stainless steel shell flushed with nitrogen. The nitrogen avoid the presence of air containing $\mathrm{^{222}Rn}$ around the detector.
%\subsubsection{The Polyethylene Shield}
%\label{sec:PE}
The lead shield is installed inside an octagonal neutron shield made of 40 cm thick high density polyethylene (HDPE) walls. This last layer allows mitigation of background induced by neutron. 

\subsection{The Gas Supply System}
\label{sec:Gas}
\begin{figure}[h!]
\centering
\includegraphics[width=120mm  ]{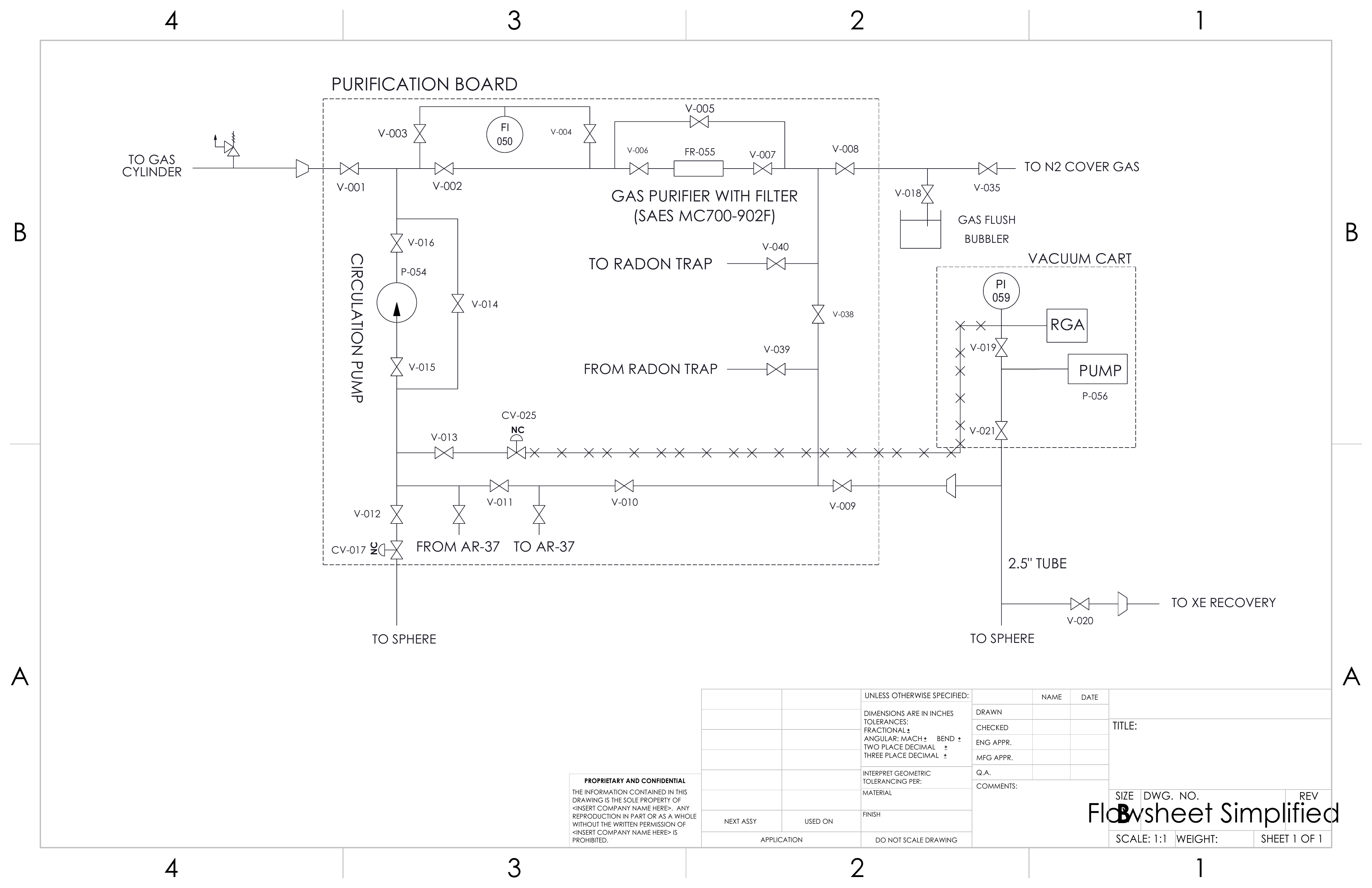}
\caption{Schematic of the gas handling system. The system allows for circulation, purification, calibration and gas quality control.}
\label{fig:GasBoard}
\end{figure}

The gas supply system is composed of three primary sub-systems: the gas bottles, the purification board, and the vacuum cart, as shown on Fig~\ref{fig:GasBoard}. %Currently, the system supports two separate gas bottles and hardware, and connects directly to the purification board. 
The purification board contains attachments for multiple sub-systems, including a gas purifier, radon trap, and circulation pump. %The detector can be operated in two modes; with circulation where the gas in constantly purified and in close mode, without circulation, the gas is then purified once when it is injected. 
It also includes connections for an $\mathrm{^{37}Ar}$ source for detector calibration~\cite{Kelly:Ar37}. The purification board is connected to the vacuum pump and a residual gas analyzer (RGA). All these components connect the detector through a glove-box (see Fig~\ref{fig:glove-box}) mounted on the HDPE shield. Every component can be isolated depending on the operating mode of the detector. Additional connection points are built into the system to allow for adding more gas cylinders.% and a future xenon recovery system. 
A separate cover gas system is in place to flush radon-free nitrogen in the space surrounding the copper sphere, as well as in the glove box and rod container. 

\newpage

\subsection{The Sensor Support Structure}
\label{sec:Rod}
\begin{figure}[h!]
\centering
\includegraphics[width=0.1\linewidth]{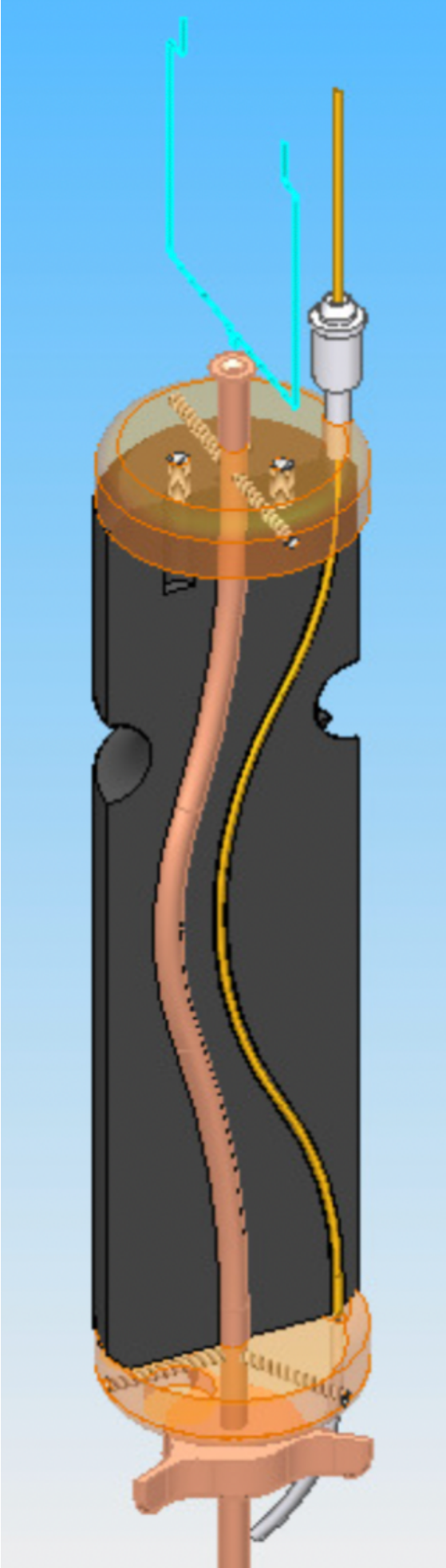}
\caption{The rod alignment system is made of a Roman lead cylinder (grey) and two copper centering disks. The lead cylinder is crossed by two curved holes for high voltage cables and the optical fiber.}
\label{fig:RodAlign}
\end{figure}

The rod is a 6\,mm diameter pipe made of C10100 copper, holding the sensor in a central position inside the sphere. Fig~\ref{fig:RodAlign} shows the alignment system of the rod. A set of alignment disks, and a cylinder of Roman lead, guarantee rod alignment within the 63\,mm diameter and 280\,mm long nozzle of the sphere. The sensor is estimated to be within 5\,mm of the center of the sphere. The 6\,mm pipe has two fixed copper disks separated by 230 mm to ensure a small angular deviation. %Inserting the rod with such disks into the sphere through the top tube, the angular deviation is certainly below the allowed deviation.
A Roman lead cylinder is inserted between the two disks to avoid any hole in the lead shielding. The lead cylinder is crossed by two curved holes for high voltage cables and the optical fiber for UV laser calibration. 

\subsection{Low and High Voltage System}
\label{sec:HV}
The detector requires low voltages, positive and negative, for the preamplifier and the digitizer, as well as high voltages, up to 5000\,V, positive and negative, for the anodes and the bias electrode. %In order for the detector to reach its full sensitivity, care must be taken to not bring any unwanted noise to the data. 
The low voltage is provided by a Wiener MPOD module with a high precision module MPV 8016H and the high voltage by a Iseg EHS 44 100x-K02 module.
%\subsection{The source deployment system and calibration}

% ************************************** %

% ************************************** %
\section{Background Mitigation}
\label{sec:BKGs}
% ************************************** %

\subsection{Electroplating}
\begin{figure}[h!]
\centering
\includegraphics[width=0.7\linewidth]{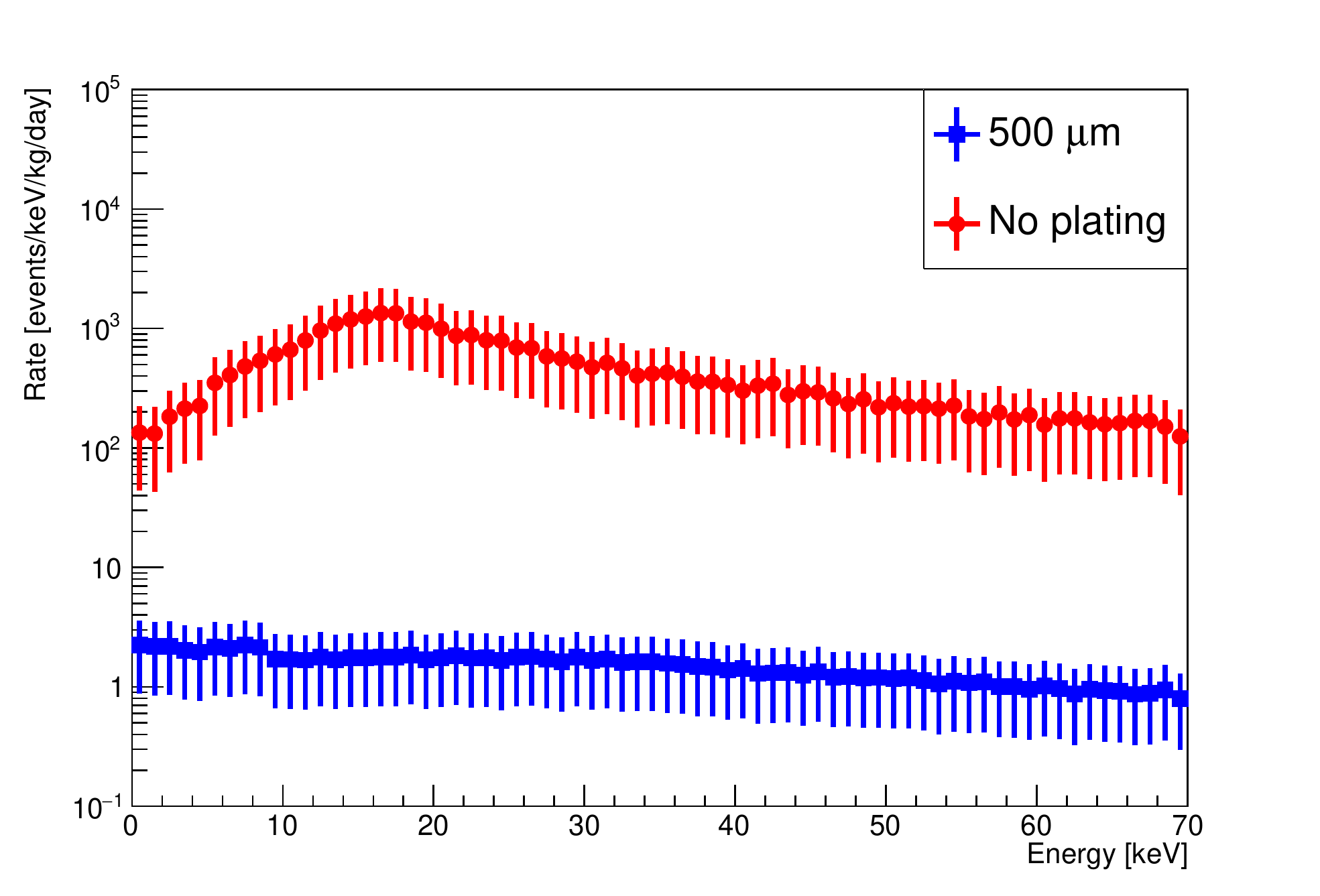}
\caption[Effect of electroplating pure copper on the inner surface of the detector.]{Effect of the electroplating of 500\,$\mathrm{\mu m}$ of pure copper on the inner surface of the detector.}
\label{fig:Plating_Decay}
\end{figure}

The $^{210}$Pb contamination in NEWS-G’s copper has
been estimated in collaboration with the XMASS~Collaboration 
to be $29^{+8 +9}_{-8 -3}$\,$\mathrm{mBq/kg}$~\cite{Balogh:2020nmo} using a low-background alpha-particle counter~\cite{Abe:2017jzw}.
Geant4~\cite{GEANT4:2002zbu} simulations of the background showed that this would represent the dominant contribution.
Thus, it was decided that a 500\,$\mathrm{\mu m}$ thick ultra-pure copper layer would be electroplated on the inner surface of the two hemispheres before welding them to form the sphere. Figure~\ref{fig:Plating_Decay} shows the results from a Geant4 simulation of the effect of such a layer of pure copper on the inner surface of the spherical vessel. 

The method used to deposit ultra-radiopure copper is potentiostatic electroforming. This process exploits the electrochemical properties of copper and contaminant species, such as lead, uranium, and thorium, to suppress the deposition of contaminants on the detector surface. The method has been developed and refined by the Pacific Northwest National Laboratory (PNNL) \cite{Hoppe2009JRADIOANALNUCLCHEM,Bunker:2020sxw}, and has been employed by rare-event search experiments, such as the Majorana Demonstrator~\cite{Majorana:2013cem}.

Radioassay studies of electroformed copper using the most sensitive assay techniques, namely Inductively Coupled Plasma Mass Spectroscopy (ICP-MS)~\cite{LaFerriere2015ANA} for $\mathrm{^{238}U}$ and $\mathrm{^{232}Th}$ and alpha-particle counting with an XIA UltraLow-1800 counter for $^{210}$Po, %and $^{210}$Pb
have demonstrated contaminations levels of less than $0.099\;\mu \mathrm{Bq/kg}$, $0.119\;\mu \mathrm{Bq/kg}$ and $5.3\;\mathrm{mBq/kg}$ for $\mathrm{^{238}U}$, $\mathrm{^{232}Th}$ and $^{210}$Po, respectively, each limited by the sensitivity of the assay method \cite{Abe:2017jzw,Abgrall:2016cct}. A complete discussion of the NEWS-G electroforming, performed at the LSM, may be found in Ref.~\cite{Balogh:2020nmo}.

%The $\mathrm{^{210}Pb}$ contamination in several C1020 commercial copper samples has been estimated by the XMASS collaboration to be in the range of 17-40 $\mathrm{mBq/kg}$ using a low-background alpha particle counter ~\cite{Abe:2017jzw}. A sample of the copper used to make the sphere was measured with this method and shows an activity of $\mathrm{28.5\pm 8.1}$ $\mathrm{mBq/kg}$~\cite{Balogh:2020nmo}. $\mathrm{^{210}Pb}$ decay is followed by de-excitation and decay of $\mathrm{^{210}Bi}$. All these processes can induce low energy events by emission of gamma rays, X-ray or Auger electrons. In particular, the $\mathrm{\beta^{-}}$ decay of $\mathrm{^{210}Bi}$ produces a bremsstrahlung X-ray. Geant4 simulations show that these bremsstrahlung radiations are a significant source of background. Without mitigation, the $\mathrm{^{210}Pb}$ decay chain in the copper would be the main source of background. Figure~\ref{fig:Plating_Decay} shows the effect of having a layer of pure copper on the inner surface of the sphere.

\subsection{Air-lock Glove Box and Cover Gas}
\label{sec:Etching}
\begin{figure}[h!]
\centering
\includegraphics[width=75mm]{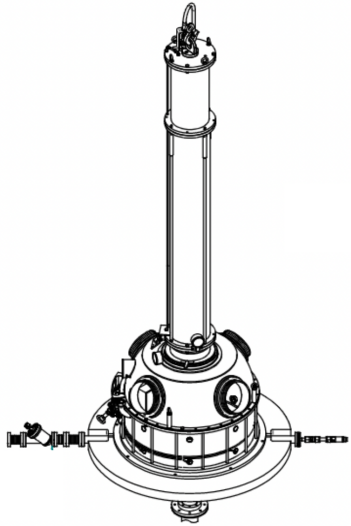}
\caption{Glove box used to install the rod and sensor. The box is filled with nitrogen before opening the sphere and lifting or lowering the rod or lowering the rod in the tube.}
\label{fig:glove-box}
\end{figure}

The air in SNOLAB contains $\mathrm{123.2 \pm 13.0\, Bq/m^{3}}$ of $\mathrm{^{222}Rn}$~\cite{SNOLAB}, and various organic impurities containing $\mathrm{^{14}C}$. $\mathrm{^{14}C}$ and $\mathrm{^{210}Pb}$, a daughter element of $\mathrm{^{222}Rn}$, can be deposited on the inner surface of the detector, and due to their half-lives being much longer than the life-time of the experiment, could be significant sources of background. 
%Although the $\mathrm{^{14}C}$ and $\mathrm{^{210}Pb}$ decay chain produce surface events that could be suppressed by a rise time cut, such backgrounds could strongly reduce the performances of the detector. 
For this reason, the inner surface of the sphere should never be exposed to air to avoid any new deposition of $\mathrm{^{14}C}$ or $\mathrm{^{210}Pb}$.  
Figure~\ref{fig:glove-box} shows the air-lock glove box and the cylinder installed on top of the detector for handling the sensor without introducing air in the detector. The glove box and cylinder are filled with nitrogen during manipulation of the sensor.  

%Sliding disks are added in the tube on top of the glove box to help keeping the rod right in the middle of the tube when being manipulated and avoid sensor damage. 
%, at the moment when only the sensor is inside and should not touch the edge of this tube. 
In order to assemble or modify the full rod mechanics, a maintenance stand has been created beside the experiment. Since the top of the rod can be precisely positioned on the top of the stand, a reference is included at the bottom of this stand "where the sensor has to be placed" ensuring a precise radial and vertical position. %by respect to the top of the north tube.
Nitrogen is continuously flushed in the stainless steel shell surrounding the lead shielding to remove air and $\mathrm{^{222}Rn}$ near the copper sphere.
%In addition, the 2 cm thick volume between the sphere and the lead shielding is filled with nitrogen to avoid background from $\mathrm{^{222}Rn}$ decay chain. 

\subsection{Gas Purification}
\label{sec:Purification}
The gases selected for the experiment are high purity 6N with impurity concentrations lower than 0.5 ppm of O${_2}$ and H${_2}$O. The gas handling system has several components for purifying the gas, such as the SAES MC1-905 purifier and a radon trap to remove electronegative impurities and $\mathrm{^{222}Rn}$, respectively. Electronegative impurities, such as water or oxygen, reduce the number of ionization electrons by electron capture and compromise the signal energy reconstruction. Therefore, the gas is circulated and continuously purified to reduce the impurities to the 10$^{-9}$ O${_2}$-equivalent level (ppb). Magboltz~\cite{Magboltz} simulations show negligible effect on the resolution at such level of contamination. The detector can also be operated without circulation in which case the gas is filtered once when injected into the sphere. 

It has been found that materials used for filtering emanate $\mathrm{^{222}Rn}$ and the deposition of its daughters elements on the inner surface of the detector is a source of low energy events. From Geant4 simulations, it has been estimated that an event rate of 1 event/kg/day for sub-keV events is reached with a $\mathrm{^{222}Rn}$ activity of 32\,$\mathrm{\mu Bq}$ for a detector filled with 135\,mbar of pure $\mathrm{CH_{4}}$. In order to reduce the contamination in the gas mixture below 32 $\mathrm{\mu Bq}$, a radon filtration system was built and integrated into the purification loop in series with the gas purifier. Radon is retained in a radon trap operated at dry ice temperatures (-80$^{\circ}$C) until it decays, while the purified Ne/CH${_4}$ based mixture is transferred into the detector. The material used for filtering the radon is a carbon molecular sieve (Carboxen 1000). It is selected for its high purity, compared to common active charcoals, in order to avoid any emanation from the trap itself.

\section{Activity Measurements and Background Simulation}
\label{sec:Activity}
% ************************************** %
\subsection{Copper Activity}
\label{sec:CuBKG}
\subsubsection{\textsuperscript{238}U and \textsuperscript{232}Th Decay Chain}
Uranium and thorium are present in the Earth's crust as a relic from its formation. Due to their long half-lives of 14.1 Gyr for $\mathrm{^{232}Th}$ and 4.5 Gyr for $\mathrm{^{238}U}$, they are currently present in rock even 4.5 Gyr after the formation of the solar system. Both $\mathrm{^{232}Th}$  and $\mathrm{^{238}U}$ decay by $\alpha$-emission and are the beginning of long decay chains. The half-lives of all daughter isotopes are much shorter than the Gyr range of $\mathrm{^{232}Th}$ and $\mathrm{^{238}U}$, therefore we assumed these decay chains are in secular equilibrium. This is the case for the $\mathrm{^{232}Th}$ decay chain, both in the copper of the vessel and the lead surrounding it. However, it is known that this is not true for $\mathrm{^{238}U}$, where the equilibrium is broken at $\mathrm{^{210}Pb}$. Moreover, in both cases a break in secular equilibrium is sometimes observed at the level of long living radium isotopes, whenever the material has gone through melting or purification, radium being chemically quite active~\cite{LZ:2020fty}. For sake of simplicity, though, in this section, we will consider the whole decay chain of $\mathrm{^{232}Th}$ and the decay chain of $\mathrm{^{238}U}$ between $\mathrm{^{238}U}$ and $\mathrm{^{210}Pb}$ to be in equilibrium, leaving more accurate studies to a future background budget paper.

%The half-lives of all daughter isotopes are much shorter than the Gyr range of $\mathrm{^{232}Th}$ and  $\mathrm{^{238}U}$, we assumed these decay chains are in secular equilibrium. This is the case for the $\mathrm{^{232}Th}$ decay chain, both in the copper of the vessel and the lead surrounding it. However, it is known that this is not true for $\mathrm{^{238}U}$, where the equilibrium is broken at $\mathrm{^{210}Pb}$. So in this section, we will consider the whole decay chain of $\mathrm{^{232}Th}$ and the decay chain of $\mathrm{^{238}U}$ between $\mathrm{^{238}U}$ and $\mathrm{^{210}Pb}$ to be in equilibrium. 

The amount of $\mathrm{^{238}U}$ and $\mathrm{^{232}Th}$ in the copper used for the NEWS-G detector was measured at PNNL using an ICP-MS. Eleven samples of copper were cut from a test hemisphere using a high-pressure water jet. The samples were cleaned to remove any oils or lubricants and etched with nitric acid to expose a virgin copper surface, rinsed in de-mineralized water, and then were fully dissolved in nitric acid. Finally, the samples were analyzed on an Agilent 8800s ICP-MS and quantified using an isotope dilution mass spectrometric method. The measured activities of $\mathrm{^{232}Th}$ and  $\mathrm{^{238}U}$ are summarized in Table~\ref{fig:TableUTh}.

\begin{table}[h!]
\centering
\begin{tabular}{cc|c|c|}
\cline{3-4}
                                   &                      & $\mathrm{^{232}Th}$ & $\mathrm{^{238}U}$ \\ \hline
\multicolumn{1}{|c|}{Sample ID}    & Cu dissolved {[}g{]} & $\mathrm{\mu Bq/kg}$      & $\mathrm{\mu Bq/kg}$     \\ \hline
\multicolumn{1}{|c|}{1}  & 2.01               & 7.7        $\mathrm{\pm}$           0.2             & 2.0         $\mathrm{\pm}$        0.1            \\ \hline
\multicolumn{1}{|c|}{2}  & 1.66               & 27.5       $\mathrm{\pm}$            0.3              & 5.6        $\mathrm{\pm}$           0.2            \\ \hline
\multicolumn{1}{|c|}{3}  & 2.28               & 7.3        $\mathrm{\pm}$           0.1             & 2.0        $\mathrm{\pm}$           0.2            \\ \hline
\multicolumn{1}{|c|}{4}  & 1.96               & 12.8       $\mathrm{\pm}$            0.1              & 2.7      $\mathrm{\pm}$             0.1            \\ \hline
\multicolumn{1}{|c|}{5}  & 2.98               & 12.5        $\mathrm{\pm}$           0.1              & 2.9      $\mathrm{\pm}$             0.03            \\ \hline
\multicolumn{1}{|c|}{6}  & 1.89               & 7.1          $\mathrm{\pm}$         0.1             & 1.8          $\mathrm{\pm}$         0.1            \\ \hline
\multicolumn{1}{|c|}{7}  & 4.58               & 22.1        $\mathrm{\pm}$           0.6              & 4.9      $\mathrm{\pm}$             0.2            \\ \hline
\multicolumn{1}{|c|}{8}  & 3.73               & 10.0        $\mathrm{\pm}$           0.1              & 2.3     $\mathrm{\pm}$              0.1            \\ \hline
\multicolumn{1}{|c|}{9}  & 5.81               & 18.3        $\mathrm{\pm}$           0.1              & 4.0      $\mathrm{\pm}$             0.1            \\ \hline
\multicolumn{1}{|c|}{10} & 3.68               & 13.3        $\mathrm{\pm}$           0.2              & 3.6       $\mathrm{\pm}$            0.1            \\ \hline
\multicolumn{1}{|c|}{11} & 3.65               & 3.77        $\mathrm{\pm}$           0.04             & 1.1      $\mathrm{\pm}$             0.1            \\ \hline
\multicolumn{2}{|c|}{Weighted mean}                                & 7.8         $\mathrm{\pm}$            0.03              & 2.8      $\mathrm{\pm}$                0.02             \\ \hline
\end{tabular}
\caption[Measure of $\mathrm{^{232}Th}$ and $\mathrm{^{238}U}$ activity in 11 copper samples]{ICPMS measurement of $\mathrm{^{232}Th}$ and $\mathrm{^{238}U}$ activity in 11 copper samples.}
\label{fig:TableUTh}
\end{table}

%\subsubsection{\bold{$\mathrm{\bold{^{40}K}}$}}
\subsubsection{\textsuperscript{40}K}
C10100 copper is a commercial component used by several experiments. The NEXT-100 experiment~\cite{NEXT:2011eyk}, a double beta decay search experiment, investigated the radiopurity  of C10100 copper and found a %contamination of 61 and 91 $\mathrm{\mu Bq/kg}$~\cite{Alvarez:2012as}.
maximal contamination of 91\,$\mathrm{\mu Bq/kg}$ of $\mathrm{^{40}K}$~\cite{Alvarez:2012as}.

%\subsubsection{$\mathrm{^{210}Pb}$ decay chain}

%$\mathrm{^{210}Pb}$ is a long-lived isotope ($\mathrm{t_{1/2}=22.2}$ years) which is in the $\mathrm{^{238}U}$ decay chain. Although the contamination of $\mathrm{^{238}U}$ is expected in the range of few tens of $\mathrm{\mu Bq/kg}$ the equilibrium is highly broken at the level of $\mathrm{^{210}Pb}$. The $\mathrm{^{210}Pb}$ contamination in commercial copper has been estimated by the XMASS collaboration to be in the range of 17-40 $\mathrm{mBq/kg}$ using a low-background alpha particle counter ~\cite{Abe:2017jzw}. A sample of the copper used to make the sphere was measured with this method and shows an activity of $\mathrm{28.5\pm 8.1}$ $\mathrm{mBq/kg}$. $\mathrm{^{210}Pb}$ decay is followed by de-excitation and decay of $\mathrm{^{210}Bi}$. All these processes can induce low energy events by emission of gamma rays, X-ray or Auger electrons. In particular, the $\mathrm{\beta^{-}}$ decay of $\mathrm{^{210}Bi}$ produces a bremsstrahlung X-ray. Figure~\ref{fig:Position_Decay} shows the radial position of the decays within the copper thickness. As expected, due to their short range in copper the electron events originate within the few first hundred $\mathrm{\mu m}$ and gamma events are rapidly attenuated but can originate from the whole thickness of copper.

%\begin{figure}[h]
%\centering
%\includegraphics[width=150mm]{Position_Decay.png}
%\caption[Position of the decays within the thickness of the copper sphere.]{Position of the decays with the thickness of the copper sphere.}
%\label{fig:Position_Decay}
%\end{figure}

\subsubsection{Cosmogenic Activation}
The main radioactive isotopes produced by cosmogenic activation in copper are $\mathrm{^{56}Co}$, $\mathrm{^{57}Co}$, $\mathrm{^{58}Co}$, $\mathrm{^{60}Co}$, and $\mathrm{^{54}Mn}$. The following table presents the half-lives and production rates of these isotopes in natural copper for the cosmogenic neutron flux at sea level. Two models of neutron flux were studied, Ziegler~\cite{5389314} and Gordon et al.~\cite{1369506}, and the most conservative results from Ziegler are quoted~\cite{Cebrian:2010zz}.

\begin{figure}[h]
\centering
\includegraphics[width=0.7\linewidth]{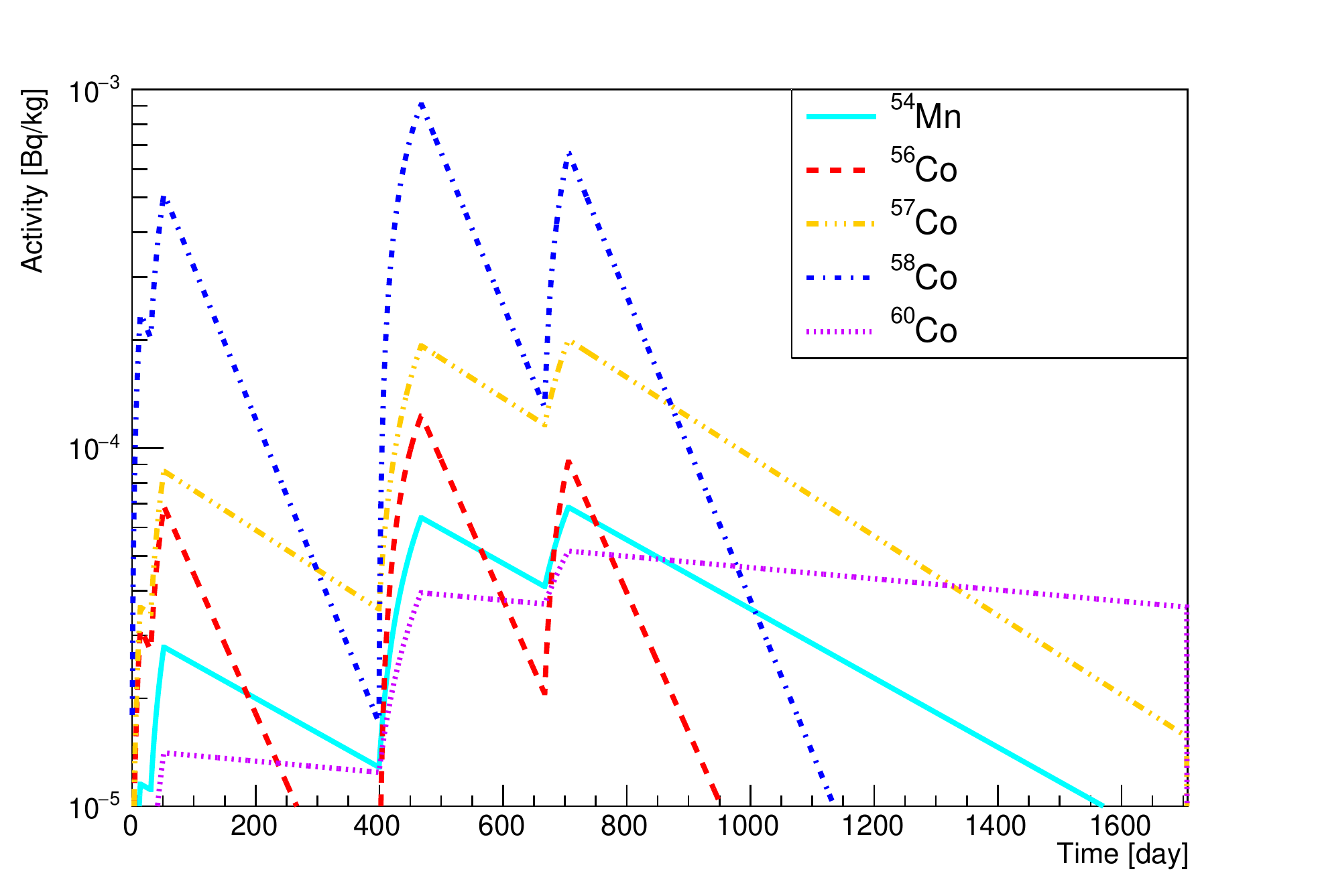}
\caption[]{Activity of the isotopes formed by cosmogenic activation during different steps of fabrication of the copper sphere.}
\label{fig:CosmoActivation.png}
\end{figure}

\begin{table}[h]
\centering
\begin{tabular}{|c|c|c|}
\hline
Isotope  & Half-life {[}days{]} & Production rate
{[}atoms/kg/day{]} \\ \hline
%$\left[\mathrm{atoms/days^{-1}}\right]$ \\ \hline
{$\mathrm{^{56}Co}$} & 77.2                 & 22.9                                                   \\ \hline
{$\mathrm{^{57}Co}$} & 271.7                & 88.3                                                   \\ \hline
{$\mathrm{^{58}Co}$} & 70.9                 & 159.6                                                    \\ \hline
{$\mathrm{^{60}Co}$} & 1898                 & 97.4                                                    \\ \hline
{$\mathrm{^{54}Mn}$} & 312                  & 32.5                                                    \\ \hline
%\multicolumn{1}{|l|}{$\mathrm{^{59}Fe}$} & 44.6                 & 6.5                                                      \\ \hline
%\multicolumn{1}{|l|}{$\mathrm{^{46}Sc}$} & 83.79                & 3.8                                                      \\ \hline
\end{tabular}
\caption[Half-life and production rate of isotopes in natural copper.]{Half-life and production rate of isotopes in natural copper~\cite{Cebrian:2010zz}.}
\end{table}
Due to their different half-lives and production rates, the concentrations of these isotopes evolve differently. Figure~\ref{fig:CosmoActivation.png} shows the evolution of the isotopes activities with time. Day zero is December 8, 2017 when the two circular copper plates were produced, arriving underground at the LSM 15 days later. They spent 18 days underground. The two plates were then sent to be formed into two hemispheres. They spent 38 days exposed to cosmic rays. The two plates stayed underground for 347 days. During this time, 500 µm of copper was plated on the inner surface. The two hemispheres were then welded together to form a sphere. Transportation and operation took 69 days. The sphere came back to the LSM, stayed for 181 days during which the commissioning run took place and was then sent to SNOLAB. The transportation took 39 days. The sphere finally arrived underground on December 13, 2019. During all these steps, the sphere travelled by truck or boat. The cosmogenic activation being dependent on altitude, the true values of activities could be a bit higher than estimated. 

\subsection{Lead Activity}
\label{sec:PbBKG}

The intrinsic radioactivity of lead is of particular concern. Due to the presence of $\mathrm{^{238}U}$ and its decay chain products in lead ores, the $\mathrm{^{210}Pb}$ activity is higher than 1\,Bq/kg, and can be even of the order of
10-100\,Bq/kg~\cite{Pattavina:2019pxw}. While refining raw lead ore, a $\mathrm{^{210}Pb}$ concentration process takes place. In fact, while other radioactive isotopes are segregated from the slag, $\mathrm{^{210}Pb}$ is concentrated.
For example, $\mathrm{^{210}Pb}$ contamination of 24 Bq/kg~\cite{EDELWEISS:2013wrh} has been measured in lead, which would have a catastrophic impact on the background of the experiment, bringing hundreds of events/kg/keV/day (dru).%A large amount of $\mathrm{^{210}Pb}$, such as 24 Bq/kg~\cite{EDELWEISS:2013wrh}, would have a catastrophic impact on the background of the experiment. % due to the bremsstrahlung radiation and X-rays from $\mathrm{^{210}Bi}$. 
The solution to this problem was to build the lead shield in two different shells; a first shell consisting of 3 cm thick Roman lead and a second shell 22 cm thick made of commercially available
very low radioactivity (VLA) lead. The Roman lead was found at the bottom of the sea, in a boat that sank roughly two thousands years ago. This period corresponds to approximately one hundred half-lives of $\mathrm{^{210}Pb}$ and led to a decrease of its activity by a factor $\mathrm{2^{100}}$ resulting in a material essentially free of $\mathrm{^{210}Pb}$. %This section will be divided in two sections focusing first on the shell of Roman lead and second on the shell of modern lead. 
%\subsubsection{The Roman lead shell}
In a similar manner to the copper of the sphere, the $\mathrm{^{238}U}$ and $\mathrm{^{232}Th}$ activity of the Roman lead were measured with ICP-MS at PNNL. Table~\ref{fig:TableUThARCH} summarizes the activity measured in three samples of Roman lead. The $\mathrm{^{40}K}$ activity in Roman lead is estimated to be < 1.3 mBq/kg~\cite{EDELWEISS:2013wrh}.

\begin{table}[h!]
\centering
\begin{tabular}{c|c|c|}
\cline{2-3}
&\multicolumn{1}{c|}{$\mathrm{^{232}Th}$} & \multicolumn{1}{c|}{$\mathrm{^{238}U}$} \\ \hline
\multicolumn{1}{|c|}{Sample ID}  & $\mathrm{\mu Bq/kg}$       & $\mathrm{\mu Bq/kg}$    \\ \hline
\multicolumn{1}{|c|}{1} & 13.6                  $\mathrm{\pm}$ 0.3              & 66                  $\mathrm{\pm}$ 8             \\ \hline
\multicolumn{1}{|c|}{2} & 5                   $\mathrm{\pm}$ 2              & 29                  $\mathrm{\pm}$ 6             \\ \hline
\multicolumn{1}{|c|}{3} & 8                   $\mathrm{\pm}$ 1             & 37                  $\mathrm{\pm}$ 9             \\ \hline
\multicolumn{1}{|c|}{Weighted mean}       & 13                   $\mathrm{\pm}$ 0.3              & 41                  $\mathrm{\pm}$ 4             \\ \hline
\end{tabular}
\caption[$\mathrm{^{232}Th}$ and $\mathrm{^{238}U}$ activity in 3 Roman lead samples.]{$\mathrm{^{232}Th}$ and $\mathrm{^{238}U}$ activity in 3 Roman lead samples.}
\label{fig:TableUThARCH}
\end{table}

%However, the EDELWEISS collaboration, which use the same Roman lead, measured its $\mathrm{^{210}Pb}$ activity and set an upper limit of 120 mBq/kg ~\cite{EDELWEISS:2013wrh}. This result, being an upper limit, it does not disagree with the assumption of a $\mathrm{^{210}Pb}$ free lead. However, before being melted to form the shell for NEWS-G, 
The lead bricks were stored for several years at the LSM. The air at LSM contains $\mathrm{^{222}Rn}$, so deposition of $\mathrm{^{210}Pb}$ on the surface of the bricks occurred.
For this reason, even though an intensive cleaning of the bricks was performed before melting them, a contamination of a few mBq/kg is still possible. %\subsubsection{The very low activity lead shell}
The activities of $\mathrm{^{238}U}$ and $\mathrm{^{232}Th}$ in the VLA lead were also measured with ICP-MS at PNNL. The results are shown in Table~\ref{fig:TableUThAL}. The $\mathrm{^{210}Pb}$ activity in the modern shield was measured at the LSM using a germanium counter to be
$\mathrm{4.6\pm 0.02}$\,Bq/kg. The $\mathrm{^{40}K}$ activity in VLA lead is estimated to be < 1.46  mBq/kg~\cite{Aprile:2011ru}.

\begin{table}[h!]
\centering
\begin{tabular}{c|c|c|}
\cline{2-3}
&\multicolumn{1}{c|}{$\mathrm{^{232}Th}$} & \multicolumn{1}{c|}{$\mathrm{^{238}U}$} \\ \hline
\multicolumn{1}{|c|}{Sample ID}  & $\mathrm{\mu Bq/kg}$     & $\mathrm{\mu Bq/kg}$    \\ \hline
\multicolumn{1}{|c|}{1} & 2.9                  $\mathrm{\pm}$ 0.5              & 180                  $\mathrm{\pm}$ 20             \\ \hline
\multicolumn{1}{|c|}{2} & $\mathrm{<1.6}$                                  & 35               $\mathrm{\pm}$ 12             \\ \hline
\multicolumn{1}{|c|}{3} & 23                  $\mathrm{\pm}$ 3              & 24                  $\mathrm{\pm}$ 10             \\ \hline
\multicolumn{1}{|c|}{Weighted Mean}       & 3.4                   $\mathrm{\pm}$ 0.5              & 48                  $\mathrm{\pm}$        7      \\ \hline
\end{tabular}
\caption[Measure of $\mathrm{^{232}Th}$ and $\mathrm{^{238}U}$ activity in 3 VLA lead samples]{Measure of $\mathrm{^{232}Th}$ and $\mathrm{^{238}U}$ activity in 3 VLA lead samples.}
\label{fig:TableUThAL}
\end{table}

\subsection{The Laboratory Environment}
The background from the laboratory environment is made of three components, muons, neutrons and gammas. The muon flux at SNOLAB is 0.27 $\mathrm{\mu /m^{2}/days}$~\cite{SNOLAB}. The neutron flux is estimated to be 4000 $\mathrm{n /m^{2}/days}$~\cite{SNOLAB}. The gamma flux has been estimated from the amount of $\mathrm{^{238}U}$, $\mathrm{^{232}Th}$, $\mathrm{^{40}K}$ in the norite rock, respectively 1.2 ppm, 3.3 ppm and 1.44 ppm~\cite{SNOLAB}.

%\subsection{Simulation software packages}

\subsection{Background Table Summary}
\label{sec:bkgtable}
All background sources were simulated with Geant4 to reproduce the conditions of the data taking at SNOLAB and the background due to cosmogenic activation scaled for Januray 2023. The simulations were done using the \emph{Shielding} physics list~\cite{PLIST} provided by Geant4 with a production cut set to 14 eV. The detector was filled with 135\,mbar of pure $\mathrm{CH_{4}}$, selected for its hydrogen content. Table~\ref{tab:summary} presents the main sources of backgrounds of the NEWS-G experiment. 
The quoted uncertainties include both uncertainties from radioactivity measurements and systematic uncertainties from simulations. The $\mathrm{^{210}Pb}$ decay chain in the copper remains the most important source of background, but is strongly mitigated by the electroplating.

\begin{table}[h!]
\begin{tabular}{|c|c|c|c|c|c|}
\hline
                                    & Source  & Contamination / flux & Unit & \begin{tabular}[c]{@{}c@{}}Rate \\ \textless 1keV {[}dru{]}\end{tabular} & \begin{tabular}[c]{@{}c@{}}Rate \\ in {[}1;5{]} keV {[}dru{]}\end{tabular} \\ \hline
\multirow{4}{*}{Copper Sphere}      & $\mathrm{^{210}Pb}$   & $\mathrm{29^{+8 +9}_{-8 -3}}$  & mBq/kg                  & $\mathrm{2 \pm 1}$ &  $\mathrm{2 \pm 1}$ \\ \cline{2-6} 
                                    & $\mathrm{^{238}U}$    & $\mathrm{2.8\pm 0.02}$     & $\mathrm{\mu Bq/kg}$    & $\mathrm{0.051 \pm 0.003}$ & $\mathrm{0.07 \pm 0.003}$ \\\cline{2-6} 
                                    & $\mathrm{^{232}Th}$   & $\mathrm{7.8\pm 0.03}$    & $\mathrm{\mu Bq/kg}$    & $\mathrm{0.19 \pm 0.008}$ & $\mathrm{0.24 \pm 0.01}$ \\ \cline{2-6} 
                                    & $\mathrm{^{40}K}$     & 91                      & $\mathrm{\mu Bq/kg}$    & $\mathrm{0.14 \pm 0.09}$ & $\mathrm{0.2 \pm 0.1}$  \\ \hline
\multirow{4}{*}{Roman Lead} & $\mathrm{^{210}Pb}$   & <25                     & mBq/kg                  & $\mathrm{ < 0.34 }$ & $\mathrm{< 0. 26 }$\\ \cline{2-6} 
                                    & $\mathrm{^{238}U}$    & $\mathrm{41\pm 4}$  & $\mathrm{\mu Bq/kg}$    & $\mathrm{0.44 \pm 0.02}$ & $\mathrm{0.57 \pm 0.02}$  \\ \cline{2-6} 
                                    & $\mathrm{^{232}Th}$   & $\mathrm{13\pm 0.3}$   & $\mathrm{\mu Bq/kg}$    & $\mathrm{0.17 \pm 0.01}$  & $\mathrm{0.22 \pm 0.01}$ \\ \cline{2-6} 
                                    & $\mathrm{^{40}K}$     & <1.3                    & mBq/kg            &    <1.7   & <2.1\\ \hline
\multirow{4}{*}{VLA Lead}           & $\mathrm{^{210}Pb}$   & $\mathrm{4.6\pm 0.02}$  & Bq/kg                   & $\mathrm{0.22 \pm 0.01}$ & $\mathrm{0.23 \pm 0.02}$  \\ \cline{2-6} 
                                    & $\mathrm{^{238}U}$    & $\mathrm{48\pm 7}$     & $\mathrm{\mu Bq/kg}$    & $\mathrm{0.06 \pm 0.01}$ & $\mathrm{0.09 \pm 0.02}$ \\ \cline{2-6} 
                                    & $\mathrm{^{232}Th}$   & $\mathrm{3.4\pm 0.5}$       & $\mathrm{\mu Bq/kg}$    & $\mathrm{0.007 \pm 0.001}$ & $\mathrm{0.01 \pm 0.002}$ \\ \cline{2-6} 
                                    & $\mathrm{^{40}K}$     & <1.46                   & mBq/kg                  & $\mathrm{< 0.20 }$  & $\mathrm{< 0.36 }$  \\ \hline
\multirow{3}{*}{Laboratory}             & gamma                 & 2.1                     & $\mathrm{\gamma/cm^{2}/s}$ & $\mathrm{0.087 \pm 0.004}$  &  $\mathrm{0.120 \pm 0.005}$ \\ \cline{2-6} 
                                    & neutron               & 4000                    & $\mathrm{n/m^{2}/day}$     & $\mathrm{0.210 \pm 0.004}$  & $\mathrm{0.072 \pm 0.002}$     \\ \cline{2-6} 
                                    & muon                  & 0.27                    & $\mathrm{\mu /m^{2}/day}$     &  $\mathrm{0.007 \pm 0.004}$  & $\mathrm{0.024 \pm 0.007}$ \\ \hline
\multicolumn{4}{|c|}{Total}                                         &     $\mathrm{3 \pm 1}$     &  $\mathrm{4 \pm 1}$   \\ \hline
%\multicolumn{4}{|c|}{Total + cosmogenic activation on Oct 1 2019}  &     $\mathrm{8 \pm 2}$    &  $\mathrm{10 \pm 2}$   \\ \hline
\multicolumn{4}{|c|}{Total + cosmogenic activation on Jan 1 2023}   &    $\mathrm{5 \pm 2}$     & $\mathrm{5 \pm 2}$   \\ \hline

%\multicolumn{4}{|c|}{Total}                                          &    $\mathrm{4 \pm 2}$      &  $\mathrm{4 \pm 2}$  \\ \hline
%\multicolumn{4}{|c|}{Total + cosmogenic activation on Oct 1 2019}  &   $\mathrm{9 \pm 2}$     &  $\mathrm{10 \pm 2}$  \\ \hline
%\multicolumn{4}{|c|}{Total + cosmogenic activation on Aug 1 2021}   &  $\mathrm{7 \pm 2}$        & $\mathrm{8 \pm 2}$  \\ \hline
\end{tabular}
\caption[Summary of the main background sources. The upper limits are not included in the total.]{Summary of the main background sources. The upper limits are not included in the total.}
\label{tab:summary}
\end{table}

\begin{figure}[h!]
\centering
\includegraphics[width=150mm]{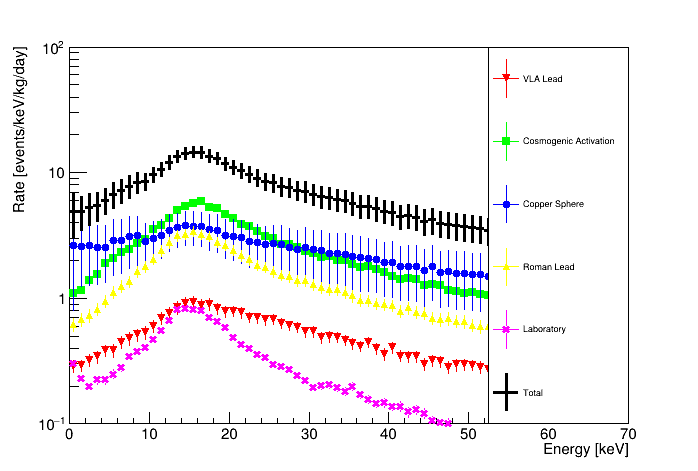}
\caption[]{Simulated energy spectrum of the NEWS-G experiment}
\label{fig:Spectrum}
\end{figure}

Figure~\ref{fig:Spectrum} shows the simulated energy spectrum of the experiment on January 1, 2023. At that time, the background is expected to be dominated by the $\mathrm{^{210}Pb}$ in the copper sphere and the cosmogenic activation of the copper sphere. The maximum observed around 16 keV is caused by Compton electrons produced within the copper. These electrons enter the gas volume with an energy of up to a few MeV, however, at these energies, their range is much longer than the detector diameter when filled with 135 mbar of pure methane. Thus, only a fraction of their energy is deposited. Data will be taken with methane and neon as target material, in addition to $\mathrm{^{222}Rn}$ that will be mitigated by filter, the possible presence of $\mathrm{^{3}H}$ in the gas is investigated and mitigation strategy will be developed if needed. %The spectrum shows a maximum around 16\,keV. This maximum is due to $\mathrm{\beta^{-}}$ coming from the inner surface of the detector. This type of events can be rejected by pulse shape discrimination. %, which can be highly suppressed with pulse shape discrimination. 
%The expected background level is roughly one order of magnitude lower than the background observed with SEDINE.

% ************************************** %
\section{Conclusion}
\label{sec:Conclusions}
% ************************************** %
The SEDINE detector, the first prototype SPC used to search for dark matter, showed the potential of such devices and set a new limit on the dark-matter-nucleon cross section. This work was a step toward the completion of a new detector installed at SNOLAB in 2021. The new detector takes advantage of the experience acquired with SEDINE. The development of the new ACHINOS sensor allows for the operation of a larger detector. The stringent selection of materials and care for background mitigation will ensure a very low background level. Study of the background shows that it is dominated by the copper sphere, with its cosmogenic activation and the presence of the $\mathrm{^{210}Pb}$ decay chain. On January 1 2023, the background level is predicted to be $\mathrm{5 \pm 2}$\,events/kg/keV/day  for sub-keV events.

%\appendix
%\section{First Appendix}
%If needed.

\acknowledgments
The help of the staff of the Laboratoire Souterrain de Modane and SNOLAB is gratefully acknowledged. This work was undertaken, in part, thanks to funding from the Canada Research Chairs program, support from the Arthur B. McDonald Canadian Astroparticle Physics Research Institute and the Canada Foundation for Innovation. This project has received support from the European Union’s Horizon 2020 research and innovation program under grant agreements No. 841261 (DarkSphere) and No. 895168 (neutronSPHERE), and from the UK Research and Innovation - Science and Technology Facilities Council through grants No. ST/S000860/1, ST/V006339/1, and ST/W005611/1.

%This is the most common positions for acknowledgments. A macro is
%available to maintain the same layout and spelling of the heading.

\newpage
\bibliographystyle{JHEP}
\bibliography{jinst-latex-sample}

\end{document}